\documentclass[epsfig,12pt]{article}
\usepackage{epsfig}
\usepackage{graphicx}
\usepackage{array}
\usepackage{color}
\usepackage{bm}
\usepackage{cite}
\usepackage{geometry}
\usepackage{latexsym}
 \usepackage{amsmath, amssymb, amscd, xypic, graphicx}

\newcommand{\beq}{\begin{equation}}   
\newcommand{\eeq}{\end{equation}}
\newcommand{\beqn}{\begin{eqnarray}}   
\newcommand{\eeqn}{\end{eqnarray}}

\newcommand*\xbar[1]{%
 \kern0.5ex%
  \hbox{%
   \kern0.2ex%
      \vbox{%
      \hrule height 0.5pt % The actual bar
      \kern0.3ex%         % Distance between bar and symbol
      \hbox{%
        \kern-0.1em%      % Shortening on the left side
        \ensuremath{#1}%
        \kern-0.1em%      % Shortening on the right side
      }%
    }%
  }%
}

\newcommand{\gsim}{\lower.7ex\hbox{$
\;\stackrel{\textstyle>}{\sim}\;$}}
\newcommand{\lsim}{\lower.7ex\hbox{$
\;\stackrel{\textstyle<}{\sim}\;$}}
\begin{document}

\begin{titlepage}

\begin{flushright}
FTPI-MINN-16/36, UMN-TH-3616/16
\end{flushright}

\vspace{5mm}

\begin{center}
{  \Large \bf  
(In)dependence of \boldmath{$\theta$}  in the  Higgs Regime\\[1mm]
 without Axions}

\vspace{6mm}
%\vspace{1mm}

 {\large   Mikhail Shifman$^{a,b}$  and  Arkady Vainshtein$^{a,b,c}$}
\end {center}

\begin{center}
$^{a}${\it  School of Physics and Astronomy, University of Minnesota,
Minneapolis, MN 55455, USA}\\[1mm]
$^b${\it  William I. Fine Theoretical Physics Institute,
University of Minnesota,
Minneapolis, MN 55455, USA}\\[1mm]
$^{c}${\it Kavli Institute for Theoretical Physics, University of California, Santa Barbara, CA 93106, USA}
\end{center}

\vspace{4mm}

\vspace{6mm}

\begin{center}
{\large\bf Abstract}
\end{center}

We revisit the issue of the vacuum angle $\theta$ dependence in weakly coupled (Higgsed) Yang-Mills theories.
Two most popular mechanisms for eliminating physical $\theta$ dependence are massless quarks and axions.
Anselm and Johansen noted that the vacuum angle  $\theta_{\rm EW}$, associated with the electroweak SU(2) in the 
Glashow-Weinberg-Salam model (Standard Model, SM), is unobservable although 
all fermion fields obtain masses through Higgsing and there is no axion.  
We generalize this idea to a broad class of Higgsed Yang-Mills theories. 
 
 In the second part we consider consequences of Grand Unification. We start from a unifying group, e.g. SU(5), at a high ultraviolet scale and evolve the theory down within the Wilson procedure. If on the way to  infrared the unifying group is broken down into a few factors, all factor groups inherit one and the same $\theta$ angle -- that of the unifying group. We show that embedding the SM in SU(5) drastically changes the
Anselm-Johansen conclusion: the electroweak vacuum angle $\theta_{\rm EW}$, equal to  $\theta_{\rm QCD}$ becomes in principle observable in $\Delta B=\Delta L =\pm 1$ processes. We also note in passing that if the axion mechanism is set up above the unification scale, we have one and the same axion in the electroweak theory and QCD, and their impacts are interdependent.

\vspace{2cm}

\end{titlepage}

%\tableofcontents
 
 %
%\begin{figure}[h]
%\epsfxsize=4cm
%\centerline{\epsfbox{mfs3}}
%\caption{\small Graphic representation for the two-dimensional sigma model with the target space (\ref{6}).}
%\label{f3}
%\end{figure}

\section{Introduction}

Shortly after introduction of the vacuum angle $\theta$ 
in QCD and in more generic Yang-Mills theories \cite{jackiw1,callan1}
it was realized that the origin of the vacuum angle and its implications 
are determined by physics at the ultraviolet (UV) scale \cite{svza}. 
Since a nonvanishing $\theta$ in QCD implies that  $CP$ is not conserved in strong interactions 
the axion mechanism of $\theta$ screening was invented
 \cite{Pec,ww}.  Another way out is a spontaneous breaking of $CP$ invariance which leads to smallness of the effective low-energy vacuum angle \cite{nelson,barr}.
 
 In the Standard Model (SM) based on the ${\rm SU}(3)\times{\rm SU}(2)\times {\rm U}(1)$ group there is another vacuum angle 
 $\theta_{\rm EW}$, associated  with  SU$(2)_{\rm weak}$\,. Although the related effects are exponentially suppressed
 the problem  was studied by Anselm and Johansen in
 \cite{anselm1} (see also \cite{perez}), with the conclusion that  $\theta_{\rm EW}$ is unobservable {\em per se}. 
 
 Here we would like to address 
 the issue of  vacuum angles and their observability in  Higgsed theories in a more general aspect. First and foremost, we assume that at low energy scales no massless fermions are left in the theory. Second, we
 do {\em not} allow for the axion mechanism to be introduced.  Under these conditions we raise the question:
 can one generalize the Anselm-Johansen mechanism making the $\theta$ angle physically unobservable?
 The answer is positive.  We prove that in a potentially broad class of Higgsed Yang-Mills theories (with no massless fermions and no axion fields) the vacuum angle $\theta$ can be absorbed in the field redefinition and is unobservable in principle. 
 
 Then we turn to the second question: $\theta$ observability
 from the standpoint of grand unification theories.
 We assume that at high energies all gauge symmetries of the model under consideration are unified in a simple group $G$.
 Then, using the Wilsonean approach we descend down to low energies where some of the
 symmetries can be spontaneously broken,  $G\to G_1\times G_2\times ...$,  (i.e. some gauge bosons become Higgsed). 
 The very fact of Grand Unification imposes constraints on the $\theta$-induced effects at low energies.
 First, above the unification scale the vacuum angle is determined by a single parameter. When we descend down in a controllable way
 this parameter does not split in two {\em independent} parameters \cite{dimopoulos2}, rather all surviving gauge groups
 $G_{1,2,..}$ inherit it from $G$. Thus, if a mechanism could be found to screen $\theta$ at short distances, then this mechanism would eliminate 
 $\theta$ in all $G_{1,2,..}$ subgroups simultaneously. 
 
 As an example we will consider SU$(5)$ unification keeping for simplicity only one generation. 
  We show that, in contradistinction with SM,  weak $\theta$ effects are in principle observable.
The  Anselm-Johansen argument \cite{anselm1} based on the analysis of the Glashow-Weinberg-Salam  model fails in SU(5). We explain why this happens. We also demonstrate that the effective $\theta$ are the same in the strong SU(3) and in the weak SU(2).

Then we make a brief comment on the general axion mechanism.
If the latter  is set up above the unification scale, we have one and the same axion in the electroweak theory and QCD, and it leads to screening vacuum angles in both sectors.
  
  We will start with a brief review of some basic points and the original Anselm-Johansen argument (see also \cite{perez,gher1}). Section \ref{gen} presents generalities. In Sec.\,\ref{31} we construct
   generalizations in  a broad class of Higgsed Yang-Mills theories. In Sec.\,\ref{41} we discuss the impact of Grand Unification. 
   In this section we argue that the $\theta$ dependence unobservable in the Glashow-Weinberg-Salam model becomes observable
   if the latter is embedded in Grand Unification. In this case instead of two independent $\theta$ angles -- one in QCD and another in the electroweak
   theory -- we deal with a single vacuum angle.  In Sec.\,\ref{41} we also make an explanatory remark about another possible source of $CP$
   violation (through Kobayashi-Maskawa matrices). 
  
  \section{Generalities: \boldmath{$\theta$} angle and its consequence}
  \label{gen}

 Let us remind that in any four-dimensional Yang-Mills theory, based on a simple group $G$ with a nontrivial homotopy
 $\pi_{3}(G)=Z$,  an additional parameter appears: the vacuum angle $\theta_{G}$. In the Lagrangian formulation
    the so called $\theta$-term is
 \beq
\Delta { \cal L}_{\theta}=\frac{\theta_{G}}{32\pi^{2}}\,{\cal F}_{\mu\nu}^{A}\widetilde{\cal F}^{\mu\nu, A}\,,
\label{theta}
\eeq
where $ {\cal F}_{\mu\nu}^A$ is the gauge field strength tensor and $A$ is the index of the adjoint representation.
Although this term is a total derivative it affects physics due to nonperturbative instanton effects.
Note that the observable nature of the vacuum angle shows up also in the Higgs regime where the theory
is at weak coupling.

Particularly, in SM the gauge group is the product ${\rm SU}(3)\times {\rm SU}(2)\times {\rm U}(1)$; hence, {\em a priori}  we deal
with two angles: $\theta_{\rm QCD}$, associated with color, and $\theta_{\rm EW}$, associated with weak SU(2). 
There are no instantons in U(1) so it does not give rise to  another angle.

While vacuum angle parameters are certainly observable and bring in $CP$ violation 
 in purely bosonic models (i.e. non-Abelian vectors plus Higgs fields and possibly scalars),
 introduction of fermion fields has a crucial impact on $\theta$-related effects. 
The fate of the latter drastically depends on fermionic matter content of the theory. 

Generically, the fermion fields are described by
a set $\psi^{a}_{\alpha}$ of the left-handed Weyl spinors plus the complex conjugated set $\bar \psi_{a,\dot \alpha}$
of right-handed spinors. The indices $\alpha,\dot \alpha=1,2$ are spinorial while the index $a$ refers to a representation of the gauge group $G$ which can be {\em reducible}. The bilinear fermionic Lagrangian takes the form
 \beq
{ \cal L}_{F}=i\bar \psi^{\dot \alpha}{\cal D}_{\alpha\dot\alpha}\psi^{\alpha}-\big[\psi^{a}_{\alpha}M_{ab}\psi^{b}_{\beta}\,\epsilon^{\alpha\beta}+{\rm H.c.}\big]
\,.
\label{lagr}
\eeq
Here ${\cal D}_{\alpha\dot\alpha}=\partial_{\alpha\dot\alpha}-i A^{A}_{\alpha\dot\alpha}T^{A}$ is the covariant derivative acting on the column of $\psi^{\alpha}$ and $M_{ab}$ is a symmetric matrix which includes possible mass terms and Yukawa couplings in the form consistent with the gauge symmetry.

In the absence of the mass terms and Yukawa couplings, $M_{ab}=0$. Addition of fermions brings in a number of U(1) symmetries associated with 
the global phase rotation for each  multiplet $\psi_{P}$,
\beq
\psi_{P} \longrightarrow {\rm e}^{\alpha_{P}}\psi_{P},
\label{rotate}
\eeq
transforming under the  the {\em irreducible} representation $P$.

We will focus on those chiral symmetries which are maintained at the classical level  but are broken by quantum anomalies.
In other words, we will consider classically conserved axial currents which are {\em not} anomaly-free. Anomalies  break their conservation. 
Moreover, nonperturbative instanton-induced effects then reveal the breaking of the anomalous global symmetries.
Famous examples include  the resolution of U(1) problem in QCD and nonconservation of baryon and lepton charges in the electroweak 
sector \cite{thooft}.

Due to these chiral anomalies the phase rotations (\ref{rotate}) lead to variations of the kinetic part $i\bar \psi^{\dot \alpha}{\cal D}_{\alpha\dot\alpha}\psi^{\alpha}$ of the fermion Lagrangian (\ref{lagr}) which are of the same form as the
$\theta$-terms (\ref{theta}), i.e. the rotations (\ref{rotate}) shift $\theta_{G}$,
\beq
\theta_{G}\longrightarrow \theta_{G}-\sum_{P} 2T(P)\,\alpha_{P}\,,
\label{hol4}
\eeq
where where $T(P)$ is the index of the representation $P$ defined as ${\rm Tr}\,(T^{a}T^{b})=T(P)\delta^{ab}$.
In the case  $M_{ab}=0$, it proves that physics is independent of the $\theta$ parameters. 
When $M_{ab}\neq 0$ there are two scenarios. If it is possible to find a combination of phase rotations which keeps the term $\psi^{a}_{\alpha}M_{ab}\psi^{b}_{\beta}\,\epsilon^{\alpha\beta}$ intact while shifting the value of the original $\theta$ as in (\ref{hol4})
 then we arrive at the same statement of $\theta$-independence as in the massless case. Otherwise, one can translate $\theta$ parameters into certain phases in $M_{ab}$ which will make physics  dependent on effective $\bar\theta$,
 \beq
 \bar \theta_{G}=\theta_{G}+{\rm arg}\,{\rm Det}\, M_{ab}\,.
 \eeq
  Below we will exemplify both scenarios  by particular examples.

\section{\boldmath{$\theta$}-independence without massless fermions\\ and axions }
\label{31}

In this section we present examples of gauge theories where the parameter $\theta$ is not observable. 
The unobservability is  due to a certain U(1) symmetry in the fermion sector. Despite this fact,  all fermions are massive in these theories.
Correspondingly, there is no need for axion in these theories. 

We start with the Anselm-Johansen construction for SU(2) case and then consider possible generalizations.

\subsection{The standard model case, SU(2)}
In this case left-handed fermions compose some number of SU(2) doublets $L^{k,A}_{\alpha}$, $k=1,2,\,\, A=1,2,\ldots$, and singlets $S^{}_{\alpha}$. Our consideration below refers to generic SU(2) case, but,
particularly, in framework of the Standard Model the first generation contains 4 doublets of SU(2)$_{\rm weak}$,
one leptonic and three quark ones,
\beq
\left(\begin{array}{c}
 \nu_\alpha\\[1mm]
 e_\alpha
 \end{array}
 \right), \quad  \left(\begin{array}{c}
u_\alpha^{\,i}\\[1mm]
d_\alpha^{\,i}
 \end{array}
 \right),\quad i=1,2,3\,,
 \label{one}
 \eeq
and 8 singlets
\beq
{\nu}^{c}_\alpha\,,\,\, {e}^{c}_\alpha\,, \,\, {u}^{c}_{\alpha,\,i}\,,\, \, {d}^{\,c}_{\alpha,\,i}\,,
 \label{two}
 \eeq
 where ${\nu}^{c}_\alpha$ is added to give (if necessary) a Dirac mass to neutrino. 
 
 The masses come from the Yukawa couplings with the standard doublet Higgs field $\phi^{i}$,
 \beq 
 {\cal L}_{\rm Y} = h_{AB}\,\bar \phi_{k}L^{k,A}_{\alpha}S^{B}_{\beta}\epsilon^{\alpha\beta}+
\tilde h_{AB}\,\epsilon_{ik}\phi^{i}L^{k,A}_{\alpha}S^{B}_{\beta}\epsilon^{\alpha\beta} +{\rm H.c.}\,\,.
 \eeq
 The phase rotations of singlets are not anomalous and can be chosen to be opposite to the rotations of the doublets.
 Keeping the Higgs field $\phi$ intact, we see that the fermion mass terms are invariant while the kinetic part produces a change of $\theta$, see Eq.\,(\ref{hol4}). It proves $\theta$-independence and the absence of  the associated $CP$ non-invariance in the SU(2) theory (with no massless fermions and no axion).
 
 Although electroweak instantons are exponentially suppressed in the electroweak SU(2) example it is interesting as a matter of principle.
 Besides it could be relevant to early cosmological time \cite{PKA}.
 
 It is also instructive to demonstrate the phenomenon in terms of the instanton calculus. The 't Hooft interaction generated by 
 antiinstanton \cite{thooft} is 
 \beq
\Delta {\cal L}_{\rm inst} \propto \prod_{A} L^{k,A}_{k}\,e^{-i\theta}\exp \Big( -\frac{8\pi^2}{g^2}\Big)
\label{four}
\eeq
where we accounted for the $\delta^{\alpha}_{k}$ structure of the fermion zero modes before averaging over the instanton SU(2) orientations.
It is clear that one can eliminate the phase factor  $\exp (-i\theta)$  by redefining  the $L$ fields. The same conclusion follows 
from the fact that only instantons contribute to processes which breaks the baryon and lepton quantum numbers, 
$\Delta B=\Delta L =\pm 1$,
in the Standard Model. Then $\exp(-i\theta)$ enters as an overall factor in the amplitude and no interference term appears for $\theta$-dependence to show up.

Thus, we see that in SM the vacuum angle $\theta_{\rm EW}$ is unobservable. This is related to the anomalous U(1) symmetry which is $B+L$ in this case, while $B-L$ is not anomalous.

\subsection{Generalization to SU{\boldmath $(N)$}}
\label{32}

Here we present a simple SU$ (N)$ generalization ($N>2$)
of the Anselm-Johansen construction. 
Consider SU$(N)$ gauge theory with $N$ complex Higgs fields $\phi^i_a$, each in the
fundamental representation of SU$(N)_{\rm gauge}$. 
Here $i=1,2,... , N$ is the color index while $a=1,2,... , N$ is the index of the flavor global SU$(N)$ symmetry.  

Prior to introduction of fermions the  model has the form
\beq
{\cal L}_B = -\frac{1}{4g^{2}}\, {\cal F}^{A}_{\mu\nu}  {\cal F}^{\mu\nu,A}+ {\cal D}^{\mu}\bar \phi^{\,a}\,{{\cal D}_\mu} \phi_{a}- V(\phi)\,,
\label{fr1}
\eeq
where $A=1,..., N^{2}-1$ is the index of the adjoint representation for field strength
tensors and summation over $A$ and $a$ is implied. The scalar sector has a global (flavor) U$(N)$ symmetry 
with the potential $V(\phi )$ chosen exactly 
as in the appropriate supersymmetric field theories (see \cite{yung}), 
\beq
V= \lambda_1 \sum_A\Big|\sum_a \bar\phi^a T^A \phi_a\Big|^2+\lambda_2 \Big|\sum_a \bar\phi^a  \phi_a - Nv^2\Big|^2 \,.
\label{fr2}
\eeq
Here $T^A$ are the generators of the color SU$(N)$, and $\lambda_{1,2}$ are constants. 
This potential can be rewritten as
\beq
V= \frac{\lambda_1}{2} \sum_{a,b}\Big[\bar\phi^{a}\phi_{b}\,\bar\phi^{b}\phi_{a} -\frac{1}{N}\,\bar\phi^{a}\phi_{a}\,\bar\phi^{b}\phi_{b}\Big]
+\lambda_2 \Big|\sum_a \bar\phi^a  \phi_a - Nv^2\Big|^2 \,.
\label{fr21}
\eeq
It is minimized  by the configuration (up to gauge transformation)
\beq
\langle \phi^{i}_{a}\rangle={\rm e}^{i\gamma}\,v\,\delta^{i}_{a}\,,
\label{fr5}
\eeq
where the phase $\gamma$ is arbitrary. 

While all $N^2\!-\!1$ gauge bosons of the original SU$(N)$ gauge theory are Higgsed \cite{yung}, acquiring one and the same 
mass $M=gv$,
one of Higgs fields remains massless representing a Goldstone boson associated with the global U(1) 
symmetry, $\phi \to \exp(i\alpha) \phi$. The arbitrariness of $\gamma$  is just a reflection of this feature. 
To make our model similar to SM we would like to eliminate the single remaining massless mode. 
This can be done in three different ways. 

First, we can 
add to the potential $V$ in Eq.\,(\ref{fr2}) a gauge invariant  term
 \beq
-\lambda_3 \,\phi^{i_1}_{a_1} \, \phi^{i_2}_{a_2} ... \, \phi^{i_N}_{a_N}\,  \varepsilon^{a_1 a_2 ... a_{N}}
 \,   \varepsilon_{i_1 i_2 ... i_{N}}+{\rm H.c.}\,.
 \label{fr6}
 \eeq
 This term, a determinant of the matrix $\phi$, is also invariant under the global flavor SU$(N)$ but it {\em breaks} the global U(1). It has dimension $N$ and is nonrenormalizable at $N>4$. 
 
 Second, we could expand the gauge sector adding a U(1) gauge boson (so that the gauge symmetry becomes
 U$(N)$ rather than SU$(N)$. Then it will be Higgsed too, and the overall $\phi$ field phase will be eaten. Needless to say, adding the  U(1) gauge will preclude us from introduction of the term (\ref{fr6}).
 
 Third, we could reduce the number of the fundamental Higgs fields from $N$ down to $N\!-\!1$.
 
 Since the essence of our construction does not change, we will demonstrate it under the assumption that the
 term (\ref{fr6}) is added to the potential, making $\gamma=0$ in Eq.\,(\ref{fr5}).

Equation (\ref{fr5}) shows that in the Higgs regime both the local gauge and global flavor groups are spontaneously
broken, but the diagonal SU$(N)$ survives as the {\em exact global} symmetry of the model. 

Now we will construct the fermion sector. The simplest vectorlike example is given by one Dirac spinor $\Psi^i $
in the fundamental representation of SU$(N)_{\rm gauge}$, which is equivalent to two left-handed Weyl spinors,
\beq
\Psi^i \longrightarrow \chi_{\alpha}^{i}\,,\,\,\, \tilde \chi^{\alpha}_{i},\qquad \alpha=1,2,\quad i=1,...,N\,,
 \label{fr14}
\eeq
fundamental and anti-fundamental with respect to SU$(N)_{\rm gauge}$ and singlets with respect to the global SU$(N)$. 
Besides there are two gauge-singlet Weyl fields which are anti-fundamental and fundamental with respect to the global SU$(N)$,
\beq
\eta^{\alpha}_{a}\,, \,\, \tilde\eta_\alpha^a \,, \qquad \alpha=1,2,\quad a=1,...,N\,.
\label{mon1}
\eeq
%The fields in (\ref{mon1}) are in the anti-fundamental representation of the global $SU(N)$ group. 

Next, we will introduce the Yukawa terms
\beq
{\cal L}_{\rm Y} =h_1 \, \chi^i \bar\phi_i^{\,a}\,\eta_a + h_2\, \tilde \chi_i \phi^i_a\,\tilde\eta^{a} +{\rm H.c.}\,.
\label{mon2}
\eeq
With the vacuum expectations value $\langle \phi^{i}_{a}\rangle=v\,\delta^{i}_{a}$ all fermions become massive; we have  $2N$ Dirac massive fermions. If $h_1 = h_2 = h$ their masses $hv$ are equal.

The Yukawa terms are invariant under the following phase rotations:
\beq 
\chi_{\alpha}^{i}\to {\rm e}^{i\alpha}\chi_{\alpha}^{i}\,,\quad \tilde\chi_{i}^{\alpha}\to {\rm e}^{i\tilde\alpha}\tilde\chi_{i}^{\alpha}\,,
\qquad \eta^{\alpha}_{a}\to {\rm e}^{-i\alpha}\eta^{\alpha}_{a}\,,\quad \tilde\eta_{\alpha}^{a}\to {\rm e}^{-i\tilde\alpha}\tilde\eta_{\alpha}^{a}\,,
 \label{mon3}
\eeq
where the gauge singlets and non-singlets are rotated in the opposite directions. The Higgs field $\phi$ stays intact. Then, the anomaly in
the rotations of non-singlet fermions leads to the shift of the vacuum angle,
\beq
\theta \longrightarrow \theta -N(\alpha +\tilde\alpha)\,.
\eeq
This makes the vacuum angle $\theta$ unobservable.

Note a crucial role of the extra U(1) given in Eq.\,(\ref{mon3})  for this conclusion. If an extra mass term of the form $m\,\chi_{\alpha}^{i}\,\tilde\chi_{i}^{\alpha}$ is added
 the $\theta$ dependence reappears.

\subsection{Quiver generalizations}

\begin{figure}[h]
\epsfxsize=7cm
\centerline{\epsfbox{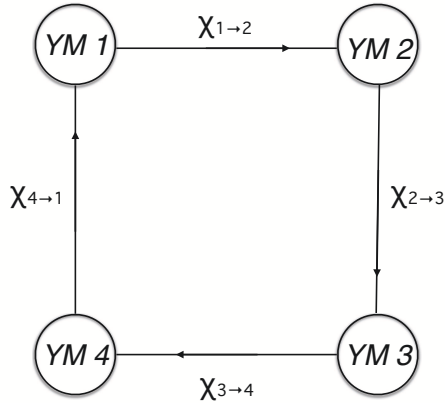}}
\caption{\small A ``square" quiver SU$(N)\times$SU$(N)\times$SU$(N)\times$SU$(N)$.  Each node represents 
a SU$(N)$ Yang-Mills theory, and each link is a bifundamental fermion, with the first index in the fundamental representation of the first SU$(N)$ while the 
second index in the anti-representation of the second SU$(N)$. For instance, on the upper horizontal link we have $\left(\chi_{1\to 2}\right)^{i_1}_{i_2}$ and so on. }
\label{fig1}
\end{figure}

It is not difficult to construct more sophisticated generalizations along these lines.
Indeed, let us consider a standard ``square" SU$(N)\times$SU$(N)\times$SU$(N)\times$SU$(N)$ quiver model depicted in 
Fig.\,\ref{fig1}.
This theory {\em per se} is chiral; hence no mass terms are possible. Generally speaking, we do not impose an 
 extra $Z_4$ symmetry, so that the gauge couplings and vacuum angles may be different in the different SU$(N)$ groups.

 Of course, in the original version of the quiver theory
all fermion fields are massless, and
$\theta$ angles are unobservable. We can make them massive through
 Higgsing similar to that discussed in Sec.\,\ref{32}.
At each node we add a scalar field with one SU$(N)_{\rm gauge}$ index and one global, similar to (\ref{fr1}).
For instance,  in the node 1 we add $\phi^{i_1}_{a_1}$ with the potential (\ref{fr2}).\footnote{The overall phase of the $\phi$ fields should be treated in the 
same way as in the previous section. } In addition, at the $1\to 2$ link we add the spinor $\eta^{i_2}_{a_1}$ with the SU$(N)_{\rm global}$ index of the given node  (i.e. 1 in the case at hand) and 
the  SU$(N)_{\rm gauge}$ index of the next  node (if we move clockwise). 
Then the gauge invariant Yukawa interaction for YM-1 has the form
\beq
h_{1} \,\chi^{i_1}_{i_2}\,\bar\phi^{\,a_1}_{i_1}\, \eta^{i_2}_{a_1}\,.
\eeq

Similar fields and interaction terms are added for YM-2, YM-3, and YM-4. Once all $\phi$ fields at each node develop expectation values (\ref{fr5}), all four SU$(N)_{\rm gauge}$ symmetries
are spontaneously broken (i.e. the corresponding gauge bosons Higgsed). This  leaves us with only 
SU$(N)_{\rm global}$ at each node. All fermion fields acquire mass terms of the type
\beq
 (h_{1} v_{1})  \,\chi^{a_1}_{i_2}\, \eta^{i_2}_{a_1} \,.
 \label{mass}
\eeq 

Despite the absence of massless quarks or axions, the $\theta$-angles are unobservable. 
Indeed,
we can rotate $\chi^{i_{1}}_{i_{2}}$ by a phase $\alpha_{1}$ 
while  $\eta^{i_{2}}_{a_{1}}$ by $-\alpha_{1}$. The mass term (\ref{mass}) obviously stays intact. On the other hand,
the vacuum angle $\theta_{1}$ of YM-1 is shifted, $\theta_{1}\to \theta_{1} - N\alpha_{1}$. Analogously,
rotations of other $\chi$ and $\eta$ fields shift the angles $\theta_{p}$ of YM-$p$. 

\section{Grand unification and {\boldmath $\theta$}-dependence}
\label{41}

\subsection{\boldmath SU$(5)$ unification}

Let us consider the simplest SU(5) unification. In this theory a  quark/lepton generation 
resides in one SU(5) decuplet $X^{[ab]}_{\alpha}$ and one antiquintet $\xbar V_{\alpha, a}$
where $a = 1,2, ... , 5$ is the SU(5) index. One weak lepton doublet $L_{\alpha}$ and three weak singlets ${d}^{\,c}_{\alpha,\,i}$ 
are form $\xbar V_{\alpha}$ while other fields from
(\ref{one}), (\ref{two})  enter $X$. 

At a large UV scale (let us call it $v_{\,\rm G}$) SU(5) is broken down to ${\rm SU}(3)\times {\rm SU}(2) \times {\rm U}(1) $ by an adjoint Higgs field. This breaking produces 12 superheavy gauge bosons with masses
$\sim gv_{\rm G}$\,. Then another Higgs field $\varphi^a$ (in the fundamental representation of SU(5)) 
breaks ${\rm SU}(3)\times {\rm SU}(2) \times {\rm U}(1)  \to {\rm SU}(3)  \times {\rm U}(1) $ at the electroweak scale $v_{{\rm EW}}$. 
Three gauge bosons acquire masses $\sim gv_{\rm EW}$. 

Responsible for the masses of the matter fields Higgs is in the fundamental representation, while the adjoint Higgs plays no role in this process. We  have the following Yukawa terms:
\beq
\lambda_1 \epsilon^{\alpha\beta}X^{[ab]}_{\alpha}\xbar V_{\beta, a}\bar\varphi_b\,, \quad \lambda_2 \epsilon^{\alpha\beta}X^{[ab]}_{\alpha}X^{[cd]}_{\beta}\varphi^{f} \varepsilon_{abcdf}\,.
\label{seven}
\eeq
 At the scale above $v_G$ 
we have a unique $\theta$-term, see (\ref{theta}). The constants $\lambda_1$ and $\lambda_2$ may have phases too. 
The effective vacuum angle $\bar\theta$, independent of redefinition of phases, is 
\beq
\bar\theta=\theta +{\rm arg}(\lambda_{1}\lambda_{2})\,.
\label{eight1}
\eeq
Indeed, in the second term one can absorb the phase of $\lambda_2$ in $\varphi$. Then, the phase of $\lambda_1$ changes, of course, ${\rm arg}\lambda_1 \to {\rm arg}\lambda_1 + {\rm arg}\lambda_2$. The latter phase can be eliminated
by a redefinition of  $\xbar V$. Simultaneously, the $\theta$-angle changes as in Eq.\,(\ref{eight1}). 
Thus, we can take $\lambda_{1,2}$ real, while $\bar\theta$ plays the role of our unique $\theta$ at the scale above $v_G$. 
%The subscript eff will be dropped hereafter. 

\vspace{2mm}

Now, let us consider the following phase rotations of the fields,
\beq
X\to X\,e^{i\alpha_{X}}\,,\quad \xbar V\to \xbar V\,e^{i\alpha_{V}}\,,\quad \varphi \to \varphi \, e^{i\alpha_{\varphi}}\,.
\label{nine}
\eeq
Due to chiral anomalies these rotations shift the $\theta$ angle by
\beq 
\theta \longrightarrow \theta -3\alpha_{X}-\alpha_{V}\,.
\label{shift}
\eeq
The coefficient 3 for the decuplet immediately follows from its content of three SU(2) doublets versus one in the quintet.
On the other hand, a simple examination of the SU(5) Yukawa terms (\ref{seven}) shows that the only classically conserved current in the
theory at hand corresponds to
\beq
\alpha_{V}=-3\alpha_{X}\,,\qquad \alpha_{\varphi}=-2\alpha_{X}\,.
\eeq
As it is seen from Eq.\,(\ref{shift}) this rotation is anomaly free and does not shift $\theta$, nor has it any impact
on the Yukawa terms (\ref{seven}).  

Thus, the parameter $\bar\theta$ {\em is in principle observable}  in the SU(5) theory.\footnote{A similar conclusion was reached by V. Rubakov, private communication, 2016.} What makes the situation different compared to
the Standard Model?
If we focus on the weak SU(2) corner of SU(5)  we immediately see the difference with the Standard Model: 
 in SU(5) there exists a perturbative amplitude with $\Delta B=\Delta L =\pm 1$ generated by Higgs particle exchanges.
The latter interferes with the instanton-induced amplitude.

Indeed, while the antiinstanton generates the vertex 
\beq
\left( X^{[ab]}X^{[cd]}X^{[fg]} V_g\, \varepsilon_{abcdf}\right)  \,e^{-i\theta}\,\exp \Big( -\frac{8\pi^2}{g^2}\Big)
\label{eight}
\eeq
(where a proper convolution of spinor indices is implied),
an amplitude with the same combination of fields  arises due to the $\varphi$ exchange through iterations of the mass terms (\ref{seven}). It has no $\theta$ and is proportional to $\lambda_{1}\lambda_{2}$. The interference of instantonic and perturbative pieces  contains a proper combination 
\beq
{\rm Re} \big(\lambda_{1}\lambda_{2}{\rm e}^{i\theta}\big)=|\lambda_{1}||\lambda_{2}|{\rm e}^{i\bar\theta}\,.
\label{eight11}
\eeq

How the $\theta$ dependence shows up at low energies? 
An appropriate framework to approach the given problem is provided by the Wilson formalism. We start from a UV scale $\mu$ which is somewhat larger than $v_{G}$, and then evolve our ``sliding scale" $\mu$ down.  In doing so we calculate the coefficients in front of various operators which appear in the Wilson Lagrangian as a 
function of $\mu$. As long as $\mu\gg v_{G}$ we can ignore symmetry breaking. Below $v_{G}$ we will have to take into account the two-step symmetry breaking, at $v_{G}$ and at $v$.  

At  $\mu\gg v_{G}$ instantons with the sizes $\rho < 1/\mu$ generate the operator  (\ref{eight}) with the running 
$g^{2}(\mu)$ and an extra coefficient 
$1/\mu^{2}$ (leaving aside a certain power of the coupling constant $g^2$). 
This is the 't Hooft interaction generated by SU$(5)$ instantons which 
is unambiguously determined. 
It was observed long ago that small-size instantons , i.e. $\rho  \ll 1/v_{G}$, probe the entire gauge group SU$(5)$
irrespective of the spontaneous breaking which may or may not occur at low energies \cite{dimopoulos2} (for a recent discussion see \cite{gher1}). 

Once $\mu$ hits $v_{G}$ the subsequent downward evolution splits in two flows: in the SU(3) and SU(2) corners, respectively. The operator (\ref{eight}) also splits in two terms, Eq. (\ref{four}) in the SU(2) corner (with $\theta_{\rm EW} = \bar\theta$) and
\beq
u\, d \, {u}^{c}{d}^{\,c}  \,e^{i\bar\theta} \exp \Big(\! -\frac{8\pi^2}{g^2}\Big) \,\,\, \mbox{in the SU(3) corner,} 
\label{eightp}
\eeq
with the color and spinor indices appropriately convoluted to form  a singlet operator. The role and fate of (\ref{eightp}) is well-known. In particular, its interference with mass terms leads to the $\theta$ dependence and CP violation at low scales.

It is worth emphasizing  that we have one and the same $\theta$  for the SU(3) and SU(2) group factors, in contradistinction to the bottom up approach of Ref.\cite{dvali1}.\footnote{\,Note, however, that in  the above publication UV completion was not considered.}  

This conclusion is valid even when other sources of $CP$ violation are present leading to a running of 
$\bar\theta$. This happens, in particular, when we deal with more than one generation. Couplings $\lambda_{1,2}$ become {\em matrices} 
with the generation indices, generally speaking 
containing $CP$ breaking phases.  Higher order iterations in these $\lambda_{1,2} $ matrices produce a logarithmic running of $\bar\theta$ similar to 
that exhibited in the calculation \cite{ellisgaillard} in SM. For several generations the definition (\ref{eight1}) of $\bar\theta$ 
must be modified in an obvious way: $\lambda_{1,2}$ must be replaced 
by determinants of these matrices.

\vspace{2mm} 

Above we explained that the $\theta$ dependence in the SU(2) corner shows up the in $\Delta B=\Delta L =\pm 1$ processes
due do interference with perturbative superheavy Higgs exchanges. It implies a suppression $m_{W}^{2}/m_{\Phi}^{2}$
for the interference besides the exponential smallness, $\exp(-8\pi^{2}/g^{2}(m_{W}))$\,. As was mentioned above, thermal enhancement of the instanton contribution in  early Universe noted in \cite{PKA} may compensate, perhaps partly, the above suppression.

\subsection{Axion} 

Now let us say a few words about the axion mechanism. The axion mechanism could be set up at the
extreme UV scale, say, by strings. Then it is natural to expect that the axion coupling at $v_{G}$
is 
\beq
-\frac{a}{32\pi^{2}f}\, {\cal F}_{\mu\nu}^A \, \tilde{{\cal F}}^{\mu\nu,\,A}
\eeq
where $ {\cal F}_{\mu\nu}^A$ is the gauge field strength tensor for SU$(5)$ and $A$ is the adjoint SU(5) index.
If so, in Eq.\,(\ref{eight}) 
\beq
\theta \to \theta -a/f\,,
\eeq
and the nonperturbative axion mass generated in the SU(3) corner makes the 
effective $\theta$ to vanish in both SU(3) and SU(2) corners {\em simultaneously}. 

\section{Conclusions}
\label{concl}

The $\theta$ angle implications in Yang-Mills theories at low energies
crucially depend on the structure of the theory in the ultraviolet. Depending on details, 
$\theta$ independence can emerge without massless quarks and/or axions. 
In this paper we prove that the Anselm-Johansen example can be generalized and, in fact, covers a broad class
of Yang-Mills theories in the Higgs regime.

 Further consequences follow from unification at a high scale.
For instance, in the SU(5) Grand Unification $\theta$-related effects in the SU(2) corner of the Glashow-Weinberg-Salam
model become observable, despite the fact that in the Glashow-Weinberg-Salam
model {\em per se} they are not. The very fact of unification implies that 
after possible spontaneous breaking $G\to G_1\times G_2\times ...$
all $G_{1,2,...}$ subgroups inherit one and the same $\bar\theta$. 

In general, if at a high scale in the ultraviolet, where ``our" physics is set up, there is a unification (with a single unifying gauge group),
and the ultraviolet analysis using no data on the subsequent evolution to low energies shows that there is no $\theta$ dependence,
this statement will remain valid at low energies although the implementation mechanisms may vary. 
Two most popular mechanisms -- massless quarks and axions -- should be supplemented in the Higgs regime 
by generalizations of the Anselm-Johansen mechanism. Relatively close scenarios are those based on 
the spontaneous breaking of $CP$ invariance. 

\section*{Acknowledgments}

We are grateful to Z. Berezhiani, G. Dvali and V. Rubakov for discussions pertinent to Sect.\,\ref{41}.

The work of M.S. was supported in part by DOE grant DE-FG02-94ER40823. 
A.V. appreciates hospitality of the Kavli Institute for Theoretical Physics where his research was supported in part by the National Science Foundation under Grant No.\ NSF PHY-1125915.

\end{document}